\documentstyle[11pt,epsfig]{article}
\date{}
\begin{document}
\title{{\bf Quadratic quantum cosmology with Schutz' perfect fluid}}
\author{Babak Vakili\thanks{%
email: b-vakili@iauc.ac.ir}\\\\
{\small {\it Department of Physics, Azad University of Chalous,}}\\{\small {\it P. O. Box 46615-397, Chalous, Iran}}}
\maketitle
\begin{abstract}
We study the classical and quantum models of a
Friedmann-Robertson-Walker (FRW) cosmology, coupled to a perfect
fluid, in the context of the $f(R)$ gravity. Using the Schutz' representation for the perfect fluid, we show that, under a particular gauge choice, it may lead to the identification of a time-parameter for the corresponding dynamical system. Moreover, this
formalism gives rise to a Schr\"{o}dinger-Wheeler-DeWitt (SWD)
equation for the quantum-mechanical description of the model under consideration, the
eigenfunctions of which can be used to construct the wavefunction of the
Universe. In the case of $f(R)=R^2$ (pure quadratic model), for some particular choices of the perfect fluid source, exact solutions to the SWD equation can be obtained and the corresponding results are compared to the usual $f(R)=R$
model. \vspace{5mm}\newline PACS numbers: 98.80.Qc, 04.50.+h,
04.60.Ds
\end{abstract}
\section{Introduction}
Over the past few years, extended theories of gravity, constructed by
adding higher-order curvature terms to the usual Einstein-Hilbert action, have
opened a new window to study the accelerated expansion of the
universe. It has been shown that such correction terms could give
rise to accelerating solutions of the field equations without having
to invoke concepts such as dark energy \cite{1}. In a more general
setting, one can use a generic function $f(R)$, instead of the usual
Ricci scalar $R$, as the action of gravitational field. In more recent times, such $f(R)$ gravity theories have been extensively studied in
the literature, see \cite{2} for a review.

In fact, quadratic and higher-order gravity theories have been used
long before the accelerated expansion of the Universe is observed, immediately after
Lovelock's natural generalization of the Einstein-Hilbert action for the gravitational field \cite{rev1}.
Quadratic Lagrangians, in particular, have been used to yield renormalizable theories of gravity coupled to matter \cite{rev2}.
They can also help us to improve the semiclassical approximation, where quantized matter fields interact with the classical gravitational field \cite{rev3}. In this context, renormalization of the energy-momentum tensor for a quantum field in a four-dimensional, curved space-time indicates that the presence of quadratic terms in the gravitational action is {\it a priori} expected \cite{rev4}. On the other hand, Kleidis {\it et al} \cite{rev5}, demonstrated that, in quadratic gravity theories, an additional coupling arises between $R^2$ and any massive quantum scalar field, introducing a {\it geometric source term} in the wave equation for the quantum field.

In order to find the dynamical equations of motion one can vary the action with respect to the
metric (metric formalism), or view the metric and connections as
independent dynamical variables and vary the action with respect to
both independently (Palatini formalism) \cite{3}. One should note
that, although in the usual Einstein-Hilbert action these two
approaches give the same field equations, in $f(R)$ gravity the
Palatini formalism leads to different dynamical equations due to
nonlinear terms in the action. However, the $f(R)$ form of the
action is shown to be equivalent to a scalar-tensor type theory and
there is an equivalence between the metric and the Palatini $f(R)$ gravity
with Brans-Dicke theories. Indeed, one can show that these versions
of $f(R)$ gravity are dynamically equivalent to the Brans-Dicke theory
of gravity with values of the Brans-Dicke parameter $\omega=0$ and $\omega=-3/2$,
respectively \cite{2}, \cite{CH}. There is also a third version of
$f(R)$ gravity in which the Lagrangian of the matter depends on the
connections of the metric (metric-affine formalism) \cite{4}.

It becomes evident that, the corrections of the $f(R)$-gravity to the results of the standard general relativity have been widely investigated, but, most of these works are restricted in the classical regime \cite{5}. The cases dealing with quantum $f(R)$ models have
seldom been studied in the literature \cite{6}, and it would be of interest
to employ such models in this study.

In this paper we shall investigate the problem of quantization of a
FRW cosmological model in the context of $f(R)$ cosmology. For the
matter field coupled to gravity we consider a perfect fluid in
Schutz' formalism \cite{7}. The advantage of using this formalism in
our quantum cosmological model is that, in a natural way, it can offer a time parameter
in terms of dynamical variables of the perfect fluid \cite{8}.
Indeed, as we shall show, after a canonical transformation the
conjugate momentum associated to one of the variables of the fluid
appears linearly in the Hamiltonian of the model. Therefore,
canonical quantization results in a Schr\"{o}dinger-Wheeler-DeWitt
(SWD) equation, in which this matter variable plays the role of time. However, it should be emphasized that we do not deal with the problem of time in the quantum cosmology of the model at hand in a fundamental way. As are well known, quantum cosmology suffers from a number of problems which one of the most important of them is the problem of time, i.e., the wave function in the Wheeler-DeWitt equation is independent of time and thus the Universe
has a static picture in this scenario. This problem was first addressed in \cite{DW} by DeWitt
himself. He argued that the problem of time should not be considered as a hindrance
in the sense that the theory itself must include a suitable well-defined time in terms
of its geometry or matter fields. Therefore, since by the Wheeler-DeWitt approach one can find different time parameters
directly from the the super-metric, the sole purpose of the use of a perfect fluid instead of a field with its own degrees of
freedom seems to be the emergence of a Schr\"{o}dinger-type equation with a time parameter $t$, which is not
the case for the general Wheeler-DeWitt equation. The Schutz model is
thus of limited value in view of the problem of time in quantum cosmology.

Here, we first consider a generic form of $f(R)$ function in the FRW
background with a generic equation of state for the perfect fluid to
construct the phase space of the model. As we find out, even for the simplest forms of the $f(R)$-function, the corresponding SWD equation is too complicated to be solved in terms of analytic functions. Therefore, we restrict ourselves to a pure fourth order model, in which $f(R) = R^2$. Accordingly, we show that, for some special forms of the equation of state of the perfect fluid, the SWD equation can be solved exactly in terms of known special functions. We then construct the resulting wavefunction of
the universe in this quadratic framework and compare its
similarities and differences with usual $f(R)=R$ model. It
is to be noted that our presentation does not claim to deal with
$f(R)$ quantum cosmology in a fundamental way since we study the
problem in a simple toy model. However, this study may reflect realistic
scenarios in other similar investigations, dealing with such problems in
a more fundamental way.
\section{The phase space of the model}
In this section we consider a FRW cosmology within the framework of
$f(R)$ gravity. In the context of the ADM formalism, the action for gravity
coupled to a perfect fluid in Schutz' representation is written as (in
what follows we work in units where $c=\hbar=16\pi G=1$)
\begin{equation}\label{A}
{\cal S}=\int_M d^4x \sqrt{-g}f(R)+2\int_{\partial M}d^3x
\sqrt{h}h_{ab}K^{ab}+\int_M d^4 x \sqrt{-g}p,\end{equation} where
$R$ is the scalar curvature and $f(R)$ is an arbitrary function of
$R$. Also, $K^{ab}$ is the extrinsic curvature and $h_{ab}$ is the
induced metric over the three dimensional spatial hypersurface,
which is the boundary $\partial M$ of the four dimensional manifold
$M$. The last term of (\ref{A}) denotes the matter contribution to
the total action where $p$ is the pressure of perfect fluid which is
linked to its energy density by the equation of state
\begin{equation}\label{B}
p=\alpha \rho.\end{equation}
In Schutz' formalism \cite{7}, the
fluid's four-velocity is expressed in terms of five potentials
$\epsilon$, $\zeta$, $\beta$, $\theta$ and $S$ as
\begin{equation}\label{C}
U_{\nu}=\frac{1}{\mu}(\epsilon_{,\nu}+\zeta \beta_{,\nu}+\theta
S_{,\nu}),\end{equation} where $\mu$ is the specific enthalpy, the
variable $S$ is the specific entropy while the potentials $\zeta$
and $\beta$ are related to torsion and are absent in the FRW models.
The variables $\epsilon$ and $\theta$ have no clear physical interpretation
in this formalism. The four-velocity satisfies the condition
\begin{equation}\label{D}
U^{\nu}U_{\nu}=-1.\end{equation}
We assume that the geometry of space–time is described by the FRW metric
\begin{equation}\label{E}
ds^2=-N^2(t)dt^2+a^2(t)\left[\frac{dr^2}{1-kr^2}+r^2(d\vartheta^2+
\sin^2 \vartheta d\varphi^2)\right],
\end{equation}
where $N(t)$ is the lapse function, $a(t)$ the scale factor and
$k$=$1$, $0$ and $-1$ correspond to the closed, flat and open
universe respectively. To proceed further, we need an effective
Lagrangian for the model whose variation with respect to its
dynamical variables yields the appropriate equations of motion.
Therefore, by considering the above action as representing a
dynamical system in which the scale factor $a$, scalar curvature $R$
and fluid's potentials play the role of independent dynamical
variables, we can rewrite the gravitational part of action (\ref{A})
as
\begin{eqnarray}\label{F}
{\cal S}_{grav}=\int dt {\cal L}_{grav}(a,\dot{a},R,\dot{R})=
\int dt \left\{Na^3f(R)-\lambda\left[R-\frac{6}{N^2}\left(\frac{\ddot{a}}{a}+
\frac{\dot{a}^2}{a^2}+\frac{k}{a^2}-\frac{\dot{N}\dot{a}}{Na}\right)\right]\right\},
\end{eqnarray}where we have introduced the definition of $R$ in terms of $a$ and its
derivatives as a constraint. This procedure allows us to remove the
second order derivatives from action (\ref{F}). The Lagrange
multiplier $\lambda$ can be obtained by variation with respect to
$R$, that is, $\lambda = Na^3 f'(R)$, in which a prime denotes the derivative with respect to $R$ . Thus, we obtain the following
point-like Lagrangian for the gravitational part of the model
\begin{equation}\label{G}
{\cal L}_{grav}=-\frac{6}{N}a\dot{a}^2f'-\frac{6}{N}a^2\dot{a}\dot{R}f''+6kNaf'+Na^3(f-Rf').\end{equation}To simplify this Lagrangian, we define the variable $\phi$ as $f'(R)=\phi$, in terms of which the Lagrangian (\ref{G}) reads
\begin{equation}\label{G1}
{\cal L}_{grav}=-\frac{6}{N}a\dot{a}^2\phi-\frac{6}{N}a^2\dot{a}\dot{\phi}+6kNa\phi-Na^3V(\phi),\end{equation}where $V(\phi)=Rf'-f=R\phi-f$.

Also, the matter part of the action (\ref{A}) becomes ${\cal S}_{matt}=\int d^3xdt Na^3p$, so the Lagrangian density of the fluid is ${\cal L}_{matt}=Na^3p$. Following the thermodynamic description of \cite{7}, the basic thermodynamic relations take the form
\begin{equation}\label{baby1}
\rho=\rho_0(1+\Pi),\hspace{0.5cm}\mu=1+\Pi+\frac{p}{\rho_0},\end{equation}where $\rho_0$ and $\Pi$ are the rest-mass density and the specific internal energy of the fluid respectively. These quantities together with the temperature of the system $\tau$, obey the first law of the thermodynamics $\tau dS=d\Pi+pd(1/\rho_0)$, where can be rewritten as
\begin{equation}\label{baby2}
\tau dS=d\Pi+pd\left(\frac{1}{\rho_0}\right)=(1+\Pi)d\left[\ln \rho_0^{-\alpha}(1+\Pi)\right],\end{equation}in which we have used the equation of state (\ref{B}). Therefore, we obtain the following expressions for the temperature and the entropy of the fluid
\begin{equation}\label{baby3}
\tau=1+\Pi,\hspace{0.5cm}S=\ln \rho_0^{-\alpha}(1+\Pi)=\ln \frac{\mu}{\alpha+1}\rho_0^{-\alpha}.\end{equation}Now, we can express $\rho_0$ and $\Pi$ as functions of $\mu$ and $S$ as
\begin{equation}\label{baby4}
1+\Pi=\frac{\mu}{\alpha+1},\hspace{0.5cm}\rho_0=\left(\frac{\mu}{\alpha+1}\right)^{1/\alpha}e^{-S/\alpha},\end{equation}so that with the help of (\ref{baby1}), one can put the equation of state in the form
\begin{equation}\label{baby5}
p=\frac{\alpha}{(\alpha+1)^{1+1/\alpha}}\mu^{1+1/\alpha}e^{-S/\alpha}.\end{equation}On the other hand, normalization of the fluid's four-velocity (\ref{C}), according to the relation (\ref{D}) implies $\mu=(\dot{\epsilon}+\theta \dot{S})/N$. Therefore, using the above constraints and  thermodynamical considerations for the fluid we find
\begin{equation}\label{H}
{\cal L}_{matt}=N^{-1/\alpha}a^3\frac{\alpha}{(\alpha+1)^{1+1/\alpha}}\left(\dot{\epsilon}+\theta \dot{S}\right)^{1+1/\alpha}e^{-S/\alpha}.\end{equation}Let us now construct the Hamiltonian for our model. The momenta conjugate to each of the above variables can be obtained from the definition $P_{q}=\frac{\partial {\cal L}}{\partial \dot{q}}$. In terms of the conjugate momenta the Hamiltonian is given by
\begin{equation}\label{I}
H=H_{grav}+H_{matt}=\dot{a}P_a+\dot{\phi}P_{\phi}+\dot{\epsilon}P_{\epsilon}+\dot{S}P_S-{\cal L},\end{equation}where ${\cal L}={\cal L}_{grav}+{\cal L}_{matt}$. Noting that
\[P_{\epsilon}=\frac{\partial {\cal L}}{\partial \dot{\epsilon}}=N^{-1/\alpha}a^3\frac{(\dot{\epsilon}+\theta \dot{S})^{1/\alpha}}{(\alpha+1)^{1/\alpha}}e^{-S/\alpha},\] and \[P_S=\frac{\partial {\cal L}}{\partial \dot{S}}=N^{-1/\alpha}a^3\frac{(\dot{\epsilon}+\theta \dot{S})^{1/\alpha}}{(\alpha+1)^{1/\alpha}}\theta e^{-S/\alpha},\] expression (\ref{I}) leads
\begin{equation}\label{J}
H=N{\cal H}=N\left[-\frac{P_aP_{\phi}}{6a^2}+\frac{\phi}{6a^3}P_{\phi}^2-6ka\phi+a^3V(\phi)+a^{-3\alpha}e^SP_{\epsilon}^{\alpha+1}\right].\end{equation}Now, consider the following canonical transformation which is a generalization of the ones used in \cite{9}
\begin{equation}\label{K}
\begin{array}{cc}
T=-P_Se^{-S}P_{\epsilon}^{-(\alpha+1)}, & P_T=P_{\epsilon}^{\alpha+1}e^S, \\\\
\bar{\epsilon}=\epsilon-(\alpha+1)\frac{P_S}{P_{\epsilon}}, & \bar{P_{\epsilon}}=P_{\epsilon}.
\end{array}
\end{equation}Under this transformation Hamiltonian (\ref{J}) takes the form
\begin{equation}\label{L}
H=N{\cal
H}=N\left[-\frac{P_aP_{\phi}}{6a^2}+\frac{\phi}{6a^3}P_{\phi}^2-6ka\phi+a^3V(\phi)+\frac{P_T}{a^{3\alpha}}\right].\end{equation}We
see that the momentum $P_T$ is the only remaining canonical variable
associated with matter and appears linearly in the Hamiltonian. The
setup for constructing the phase space and writing the Lagrangian
and Hamiltonian of the model is now complete.

The classical dynamics is governed by the Hamiltonian equations, that is
\begin{eqnarray}\label{M}
\left\{
\begin{array}{ll}
\dot{a}=\{a,H\}=-\frac{N}{6}\frac{P_{\phi}}{a^2},\\\\
\dot{P_a}=\{P_a,H\}=N\left[-\frac{1}{3}a^{-3}P_aP_{\phi}+\frac{1}{2}\phi a^{-4}P_{\phi}^2+6k\phi -3a^2V(\phi)+3\alpha a^{-3\alpha-1}P_T\right],\\\\
\dot{\phi}=\{\phi,H\}=N\left[-\frac{1}{6}\frac{P_a}{a^2}+\frac{1}{3}\frac{\phi}{a^3}P_{\phi}\right],\\\\
\dot{P_{\phi}}=\{P_{\phi},H\}=N\left[-\frac{1}{6}\frac{P_{\phi}^2}{a^3}+6ka-a^3V'(\phi)\right],\\\\
\dot{T}=\{T,H\}=\frac{N}{a^{3\alpha}},\\\\
\dot{P_T}=\{P_T,H\}=0.
\end{array}
\right.
\end{eqnarray}
We also have the constraint equation ${\cal H}=0$. Up to this point
the cosmological model, in view of the concerning issue of time, has
been of course under-determined. Before trying to solve these
equations we must decide on a choice of time in the theory. The
under-determinacy problem at the classical level may be removed by
using the gauge freedom via fixing the gauge. A glance at the above
equations shows that choosing the gauge $N=a^{3\alpha}$, we have
\begin{equation}\label{N}
N=a^{3\alpha}\Rightarrow T=t,\end{equation}which means that variable $T$ may play the role of time in the model. Therefore, the classical equations of motion can be rewritten in the gauge $N=a^{3\alpha}$ as follows
\begin{eqnarray}\label{O}
\left\{
\begin{array}{ll}
\dot{a}=-\frac{1}{6}a^{3\alpha-2}P_{\phi},\\\\
\dot{P_a}=-\frac{1}{3}a^{3\alpha-3}P_a P_{\phi}+\frac{1}{2}\phi a^{3\alpha-4}P_{\phi}^2+6 k a^{3\alpha}\phi-3a^{3\alpha +2}V(\phi)+3\alpha a^{-1}P_0,\\\\
\dot{\phi}=-\frac{1}{6}a^{3\alpha-2}P_{a}+\frac{1}{3}a^{3\alpha-3}\phi P_{\phi},\\\\
\dot{P_{\phi}}=-\frac{1}{6}a^{3\alpha-3}P_{\phi}^2+6ka^{3\alpha+1}-a^{3\alpha+3}V'(\phi),
\end{array}
\right.
\end{eqnarray}where we take $P_T=P_0=\mbox{const.}$ from the last equation of (\ref{M}). Since integrability
of this system directly depends on the choice of a form for $f(R)$ (which determines potential $V(\phi)$), it is appropriate to
concentrate on this point first. However, before choosing such a function, let us deal with the quantum cosmology of the model described above.
\section{Quantization of the model}
We now focus attention on the study of the quantum cosmology of the
model described above. We start by writing the Wheeler-DeWitt
equation from Hamiltonian (\ref{L})\footnote{The canonical
transformation (\ref{K}) is applied to the classical Hamiltonian
(\ref{J}), resulting in Hamiltonian (\ref{L}) which we are going to
quantize. To make this acceptable, one should show that in the
quantum theory the two Hamiltonians are connected by some unitary
transformation, i.e. the transformation (\ref{K}) is also a quantum
canonical transformation. A quantum canonical transformation is
defined as a change of the phase space variables $(q,p)\rightarrow
(q',p')$ which preserves the Dirac bracket
\cite{10}\[\left[q,p\right]=i=\left[q'(q,p),p'(q,p)\right].\]Such a
transformation is implemented by a function $C(q, p)$ such
that\[q'(q,p)=CqC^{-1},\hspace{0.5cm}p'(q,p)=CpC^{-1}.\]This
canonical transformation $C$ transforms the Hamiltonian as
$H'(q,p)=CH(q,p)C^{-1}$. For our case the canonical relations
$[S,P_S]=[\epsilon,P_{\epsilon}]=i$
yield\[\left[T,P_T\right]=\left[-P_Se^{-S}P_{\epsilon}^{-(\alpha+1)},P_{\epsilon}^{\alpha+1}e^S\right]=-\left[P_Se^{-S},e^S\right]=
\left[e^S,P_S\right]e^{-S}=ie^Se^{-S}=i,\]and
\[\left[\bar{\epsilon},\bar{P_{\epsilon}}\right]=
\left[\epsilon-(\alpha+1)P_SP_{\epsilon}^{-1},P_{\epsilon}\right]=\left[\epsilon,P_{\epsilon}\right]=i,\]which
means that the transformation (\ref{K}) preserves the Dirac brackets
and thus is a quantum canonical transformation. Therefore, use of
the transformed Hamiltonian (\ref{L}) for quantization of the model
is quite reasonable.}. Since the lapse function $N$ appears as a
Lagrange multiplier in this Hamiltonian, we have the Hamiltonian
constraint ${\cal H}=0$. Thus, application of the Dirac quantization
procedure demands that the quantum states of the Universe should be annihilated by
the operator version of ${\cal H}$, that is
\begin{equation}\label{P}
{\cal H}\Psi(a,\phi,T)=\left[-\frac{P_aP_{\phi}}{6a^2}+\frac{\phi}{6a^3}P_{\phi}^2-6ka\phi+a^3V(\phi)+\frac{P_T}{a^{3\alpha}}\right]\Psi(a,\phi,T)=0,
\end{equation}where $\Psi(a,\phi,T)$ is the wavefunction of the universe. A remark about the reduced Hamiltonian in the above procedure is
the factor-ordering problem when one embarks on constructing a quantum mechanical
operator equation. In dealing with such Hamiltonians at the quantum level extra care should be taken when one tries to replace the dynamical variables with their quantum operator counterparts, that is , in replacing a variable $q$ and its momentum $p_q$ with their corresponding operators, the ordering
considerations should be taken into account. Therefore, to guarantee Hermiticity, the operator form corresponding to equation (\ref{P}) may be
written as
\begin{eqnarray}\label{Q}
\left[-\frac{1}{12}\left(a^rP_a a^s+a^s P_a a^r\right)P_{\phi}+\frac{1}{12a^3}\left(\phi^u P_{\phi}\phi^v P_{\phi}\phi^w+\phi^w P_{\phi}\phi^v P_{\phi}\phi^u\right)\nonumber \right.\\ \left. - 6ka \phi+a^3V(\phi)+\frac{P_T}{a^{3\alpha}}\right]\Psi(a,\phi,T)=0,\end{eqnarray}where the parameters $r$, $s$, $u$, $v$ and $w$ satisfy $r+s=-2$, $u+v+w=1$ and denote the
ambiguity in the ordering of factors $a$ and $P_a$ in the first and $\phi$ and $P_{\phi}$ in the second term
of (\ref{P}). With the replacement $P_a \rightarrow -i\frac{\partial}{\partial a}$ and similarly for $P_{\phi}$ and $P_T$ the
above equation reads
\begin{eqnarray}\label{R}
\left[\frac{1}{6}a^{-2}\frac{\partial^2}{\partial a \partial \phi}-\frac{1}{3}a^{-3}\frac{\partial}{\partial \phi}-\frac{1}{6}a^{-3}\phi \frac{\partial^2}{\partial \phi^2}+\frac{1}{6}A\phi^{-1}a^{-3}-6ka\phi \nonumber \right.\\ \left. + a^3V(\phi)-a^{-3\alpha}i\frac{\partial}{\partial T}\right]\Psi(a,\phi,T)=0,\end{eqnarray}where $A=uw$. This equation takes the form of a Schr\"{o}dinger equation $i\partial \Psi/\partial T=H\Psi$, in which the new Hamiltonian operator is Hermitian for any choice of the ordering parameters with the standard inner product
\begin{equation}\label{S}
<\Phi\mid \Psi>=\int_{(a,\phi)} a^{-3\alpha}\Phi^{\ast}\Psi da d\phi.\end{equation}
We separate the variables in the SWD equation (\ref{R}) as
\begin{equation}\label{T}
\Psi(a,\phi,T)=e^{-iET}\psi(a,\phi),\end{equation}leading to
\begin{equation}\label{U}
\left\{a\phi\frac{\partial^2}{\partial a \partial \phi}-\phi^2\frac{\partial^2}{\partial \phi^2}-2\phi\frac{\partial}{\partial \phi}+\left[
A-36ka^4\phi^2+6a^6\phi V(\phi)-6Ea^{3-3\alpha}\phi\right]\right\}\psi(a,\phi)=0,\end{equation} where $E$ is a separation constant. It is seen that this equation has a mixed derivative with respect to the variables $a$ and $\phi$ and also, these variables appear in a mixed form in the last term of the equation. Under these conditions equation (\ref{U}) cannot be solved by the method of separation of variables. Therefore, it is useful to introduce the following change of variables
\begin{equation}\label{V}
x=a\phi^{1/2},\hspace{.5cm}y=\phi.\end{equation}In terms of these variables equation (\ref{U}) takes the form
\begin{eqnarray}\label{W}
\left\{\frac{1}{4}x^2\frac{\partial^2}{\partial
x^2}-\frac{1}{4}x\frac{\partial}{\partial
x}-y^2\frac{\partial^2}{\partial y^2}-2y\frac{\partial}{\partial
y}+\left[A-36kx^4+6x^6y^{-2}V(y)-6Ex^{3-3\alpha}y^{\frac{1}{2}(3\alpha-1)}\right]\right\}\psi(x,y)=0.\end{eqnarray}
Unfortunately, this equation cannot be solved analytically for an
arbitrary potential function $V(y)$ and parameter $\alpha$, which
represent the form of $f(R)$ and the nature of the perfect fluid
respectively. In the next section we shall present a class of exact
solutions for this equation in pure $R^2$ cosmology for two special
cases of perfect fluid.
\section{Quantum solutions for $f(R)=R^2$ cosmology}
For the pure fourth order ($f(R)=R^2$), model we have $\phi=f'(R)=2R$ and $V(\phi)=Rf'-f=R^2$. Hence, from (\ref{V}) we obtain $V(\phi)=\frac{1}{4}\phi^2=\frac{1}{4}y^2$. Therefore, equation (\ref{W}) reduces to
\begin{equation}\label{X}
\left\{\frac{1}{4}x^2\frac{\partial^2}{\partial x^2}-\frac{1}{4}x\frac{\partial}{\partial x}-y^2\frac{\partial^2}{\partial y^2}-2y\frac{\partial}{\partial y}+\left[A-36kx^4+\frac{3}{2}x^6-6Ex^{3-3\alpha}y^{\frac{1}{2}(3\alpha-1)}\right]\right\}\psi(x,y)=0.\end{equation}
It is seen that in the cases of $\alpha=1, 1/3$, the variables $x$ and $y$ can be separated from each other and equation (\ref{X}) may be solved analytically for these two cases. Thus, in what follows we restrict ourselves to these two special cases.
\subsection{Stiff matter: $\alpha=1$}
Stiff matter is a fluid with pressure equal to energy density and
speed of sound equal to speed of light. In this case we separate the
solutions of equation (\ref{X}) into the form $\psi(x,y)=X(x)Y(y)$
leading to
\begin{eqnarray}\label{Y}
\left\{
\begin{array}{ll}
\left[x^2\frac{d^2}{dx^2}-x\frac{d}{dx}+\left(1-\nu^2-144kx^4+6x^6\right)\right]X(x)=0,\\\\
\left[y^2\frac{d^2}{dy^2}+2y\frac{d}{dy}+\left(6Ey-\frac{\nu^2-1}{4}\right)\right]Y(y)=0,
\end{array}
\right.
\end{eqnarray}
where we take $\frac{\nu^2-1}{4}$ as a separation constant and also
$A=0$ without losing general character of the wavefunction. One
should note that there exists an infinite number of possibilities of
ordering. As Hawking and Page have shown \cite{11}, the factor
ordering parameter will not affect semiclassical calculations in
quantum cosmology and so for convenience one usually choose a
special value for it in the special models. On the other hand, in
general, the behavior of the wavefunction depends on the chosen
factor ordering \cite{12}. But in the model at hand, as is clear
from equation (\ref{X}), the factor $A$ will appear in the solutions
of (\ref{Y}) only together with the separation constant $\nu$ (for
example $\nu \rightarrow \sqrt{\nu^2-4A}$ in the first equation of
(\ref{Y})). Therefore, since we finally take a superposition over
all values of this separation constant, we expect that the essential
features of the wavefunction will remain the same for different choices of
the factor ordering parameter $A$. Also, with an eye to equation (\ref{Q}), it is to be noted that the choice $A=0$ yields a zero value for $u$ or $w$ or both of them. In all of these cases the Hermiticity of the second term in equation (\ref{Q}) will remain unchanged. From this point of view the particular choice $A=0$ does not reflect major effects to equation (\ref{Q}).

For the flat FRW metric ($k=0$), the above equations have the following solutions in terms of Bessel functions
\begin{eqnarray}\label{Z}
\left\{
\begin{array}{ll}
X(x)=x\left[c_1J_{\frac{\nu}{3}}\left(\sqrt{\frac{2}{3}}x^3\right)+c_2Y_{\frac{\nu}{3}}
\left(\sqrt{\frac{2}{3}}x^3\right)\right],\\\\
Y(y)=\frac{1}{\sqrt{y}}\left[d_1J_{\nu}\left(\sqrt{24Ey}\right)+d_2Y_{\nu}\left(\sqrt{24Ey}\right)\right],
\end{array}
\right.
\end{eqnarray}where $c_i$ and $d_i$ ($i=1,2$) are integration constants. Thus, the eigenfunctions of the SWD equation for stiff matter, can be
written as
\begin{equation}\label{AB}
\Psi_{\nu E}(x,y,T)=e^{-iET}\frac{x}{\sqrt{y}}J_{\frac{\nu}{3}}\left(\sqrt{\frac{2}{3}}x^3\right)J_{\nu}\left(\sqrt{24Ey}\right),\end{equation}
where we choose $c_2=d_2=0$ for having well-defined functions in all ranges of variables $x$ and $y$. We may now write the general solutions to
the SWD equations as a superposition of the eigenfunctions, that is
\begin{equation}\label{AC}
\Psi(x,y,T)=\int_{E=0}^{\infty}\int_{\nu=0}^\infty A(E)C(\nu)\Psi_{\nu E}(x,y,T)dE d\nu,
\end{equation}
in which $A(E)$ and $C(\nu)$ are suitable weight functions to
construct the wave packets. By using the equality \cite{13}
\begin{equation}\label{AD}
\int_0^\infty e^{-a r^2}r^{\nu+1}J_{\nu}(br)dr=\frac{b^{\nu}}{(2a)^{\nu+1}}e^{-\frac{b^2}{4a}},\end{equation}
we can evaluate the integral over $E$ in (\ref{AC}) and simple
analytical expression for this integral is found if we choose the function $A(E)$ to be
a quasi-Gaussian weight factor $A(E)=12(24E)^{\nu/2}e^{-24\gamma E}$, which results in
\begin{equation}\label{AE}
\int_0^\infty A(E)e^{-iET}J_{\nu}\left(\sqrt{24Ey}\right)dE=\frac{y^{\nu/2}}{\left(2\gamma+\frac{iT}{12}\right)^{\nu+1}}\mbox{exp}
\left(-\frac{y}{4\gamma+\frac{iT}{6}}\right),
\end{equation}where $\gamma$ is an arbitrary positive constant. Substitution of the above relation into equation (\ref{AC}) leads to the following expression for the wavefunction
\begin{equation}\label{AF}
\Psi(x,y,T)=\frac{x}{\sqrt{y}}\mbox{exp}
\left(-\frac{y}{4\gamma+\frac{iT}{6}}\right)\int_0 ^{\infty}C(\nu)\frac{y^{\nu/2}}{\left(2\gamma+\frac{iT}{12}\right)^{\nu+1}}J_{\frac{\nu}{3}}\left(\sqrt{\frac{2}{3}}x^3\right)d\nu,
\end{equation}where $C(\nu)$ can now be chosen as a shifted Gaussian weight function $e^{-a(\nu-b)^2}$. It is seen that this expression is too complicated for extracting an analytical closed form for the wavefunction and the choice of a function $C(\nu)$ that leads to an analytical solution for the wavefunction is not an easy task. In this respect, our choices for $A(E)$ and $C(\nu)$ as quasi-Gaussian and shifted Gaussian weight functions respectively, appear to have physical grounds. Such Gaussian weight functions are widely
used in quantum mechanics as a way to think about the localized states. This is because that these types of weight factors are centered about a special value of their argument and they fall off rapidly away from that center. Due to this behavior the corresponding wave packet resulting from (\ref{AF}) after integration, has also a Gaussian-like behavior, i.e., is localized about some special values of its arguments. Therefore, it is seen that there is a reciprocal relation between the width of the Gaussian weight function that determines the shape of the
wave packet, and the width of the wave packet. In this sense the motion of the peaks of the wave packets with a group velocity and also its spreading behavior (see below), can be best seen in terms of the Gaussian packets.

\begin{figure}
\begin{tabular}{c}\hspace{-1cm}\epsfig{figure=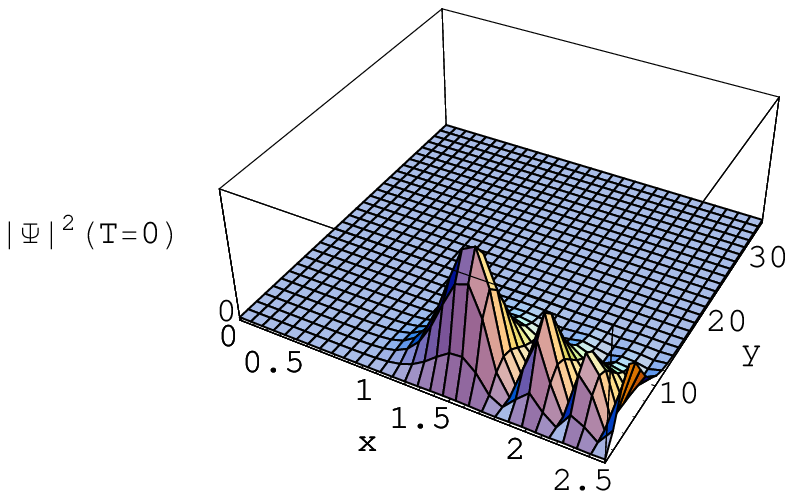,width=7cm}
\hspace{1cm} \epsfig{figure=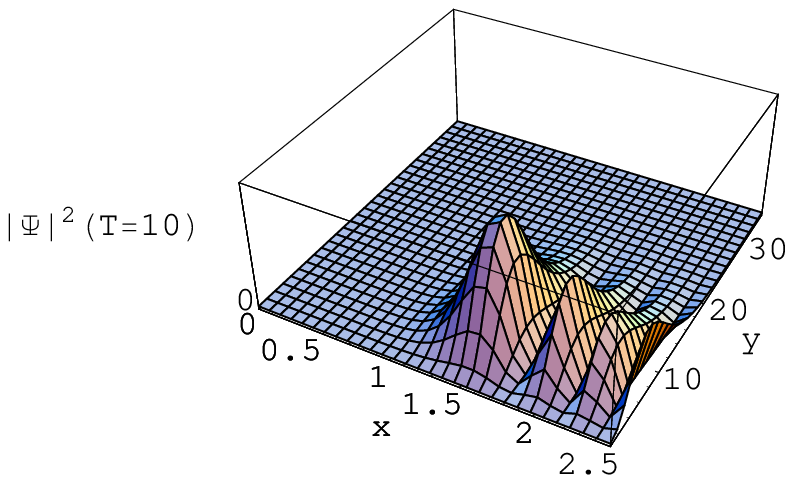,width=7cm}
\end{tabular}
\begin{tabular}{c}\hspace{-1cm}\epsfig{figure=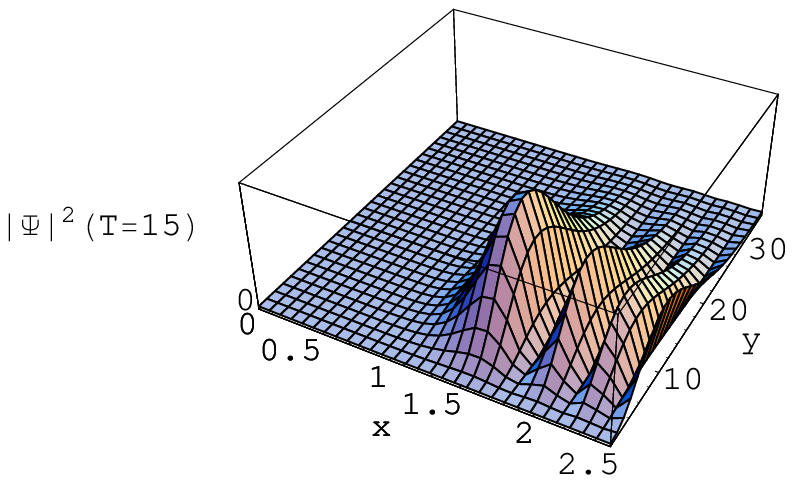,width=7cm}
\hspace{1cm} \epsfig{figure=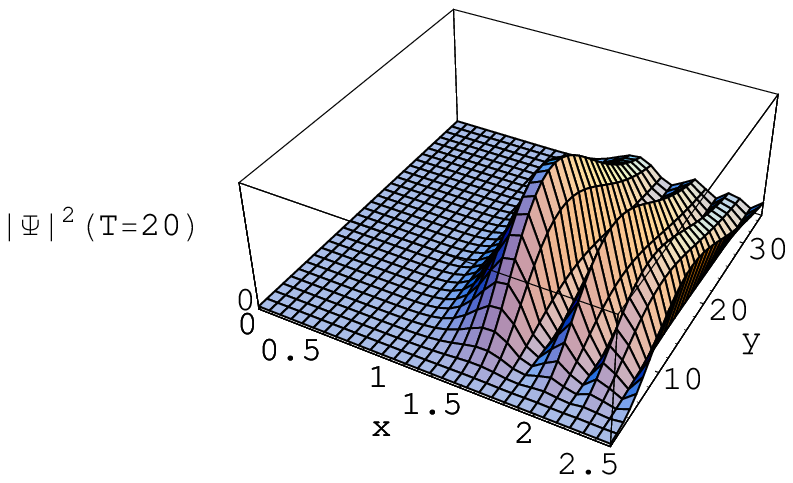,width=7cm}
\end{tabular}
\caption{\footnotesize  The figures show
$|\Psi(x,y,T)|^2$, the square of the wavefunction in four different time parameter $T=0,10,15,20$.
The figures are plotted for the numerical values $a = 2$, $b = 5$ and $\gamma = 0.5$, and we have taken
the integral in (\ref{AF}) from $0$ to $10$.}\label{fig1}
\end{figure}

In figure \ref{fig1} we have plotted the square of the
wavefunction for typical numerical values of the parameters.
As this figure shows, at $T=0$, the wavefunction has a dominant peak
in the vicinity of some nonzero values of $x$ and $y$ followed by
smaller peaks, which as $x$ grows, their amplitudes are suppressed.
This means that the wavefunction predicts the emergence of the
universe from a state corresponding to its dominant peak. However,
the emergence of several peaks in the wave packet may be interpreted
as a representation of different quantum states that may communicate
with each other through tunneling. This means that there are
different possible universes (states) from which our present
universe could have evolved and tunneled in the past, from one
universe (state) to another. As time progresses, the wave packet
begins to propagate in the $y$-direction, its width becoming wider
and its peaks moving with a group velocity towards the greater
values of $y$. Bearing in mind that $y=\phi=f'(R)$, this
wavefunction predicts that the universe will assume states with
larger $R$ in its late time evolution.

\begin{figure}
\begin{tabular}{cccc} \epsfig{figure=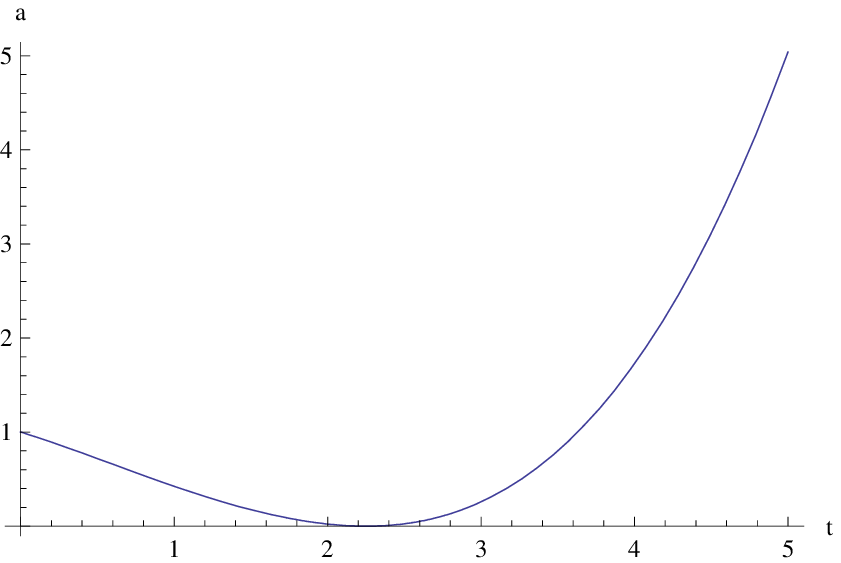,width=5cm}
\hspace{1cm} \epsfig{figure=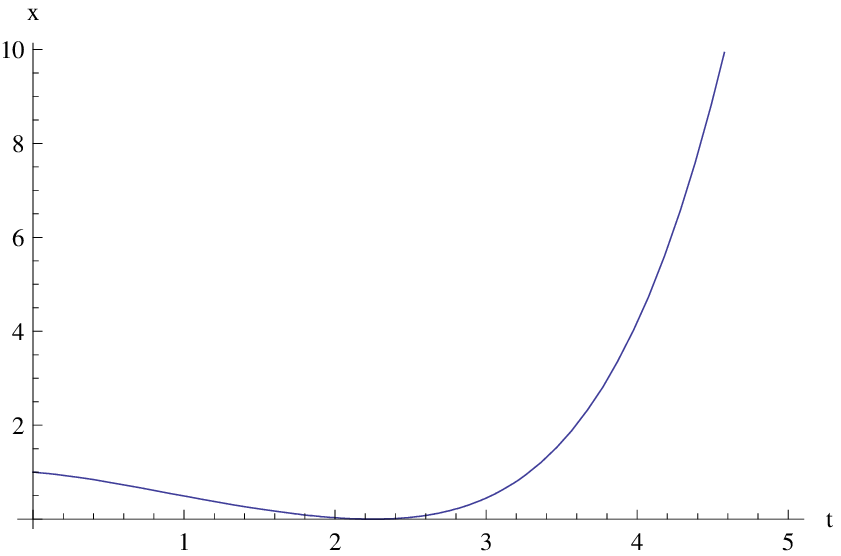,width=5cm}\hspace{1cm}\epsfig{figure=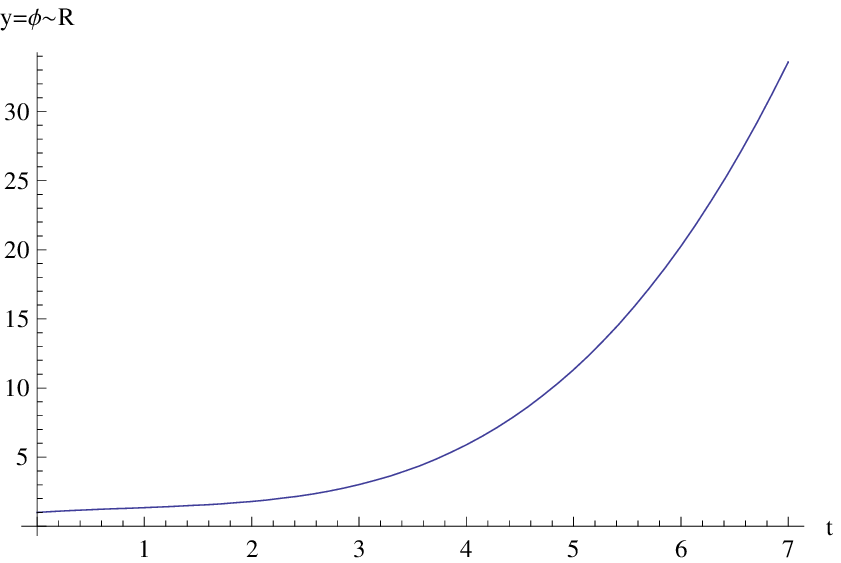,width=5cm}
\end{tabular}
\caption{\footnotesize Approximate behavior of $a(t)$, the scale factor, $x(t)=a\phi^{1/2}$ and $R(t)$, the Ricci scalar, in the classical cosmology framework. The figures are plotted for numerical values $\alpha=1$, $k= 0$ and $P_0=1$. We take the initial conditions $a(t=0)=\phi(t=0)=1$ and $P_a(t=0)=P_{\phi}(t=0)=3$. After examining other sets of initial conditions, we verify that this behavior repeats itself.}
\label{fig2}
\end{figure}

Before going any further, let us take a look at the classical
cosmology corresponding to the above quantum solutions. For the
quadratic model, we insert $V(\phi)=\frac{1}{4} \phi^2$ in the
system (\ref{O}), take $\alpha=1$ for the matter character and also
$k=0$ representing the flat FRW space time. We see that the
classical cosmology forms a system of nonlinear coupled differential
equations which unfortunately cannot be solved analytically. In
figure \ref{fig2}, employing numerical methods, we have shown the
approximate behavior of $a(t)$, $x(t)$ and $\phi(t)\sim R(t)$ for
typical values of the parameters and initial conditions
respectively.
As is clear from the figures, while the Ricci scalar has a
monotonically growing behavior with time, the scale factor first
decreases and then increases forever.
On the other hand, take a look at the figure \ref{fig1} shows that the quantum states with larger (smaller) peaks, i.e., the states of high (low) probability, are located at the smaller (larger) values for the scale factor. Therefore, when the scale factor decreases (increases) with time in the
classical point of view, in the quantum domain tunneling  occurs
from a state with a smaller (larger) peak to another with a larger
(smaller) one. Also, in view of the behavior of the Ricci scalar,
the classical and quantum solutions are in complete agreement with
each other and both predict a monotonically increasing evolution for
this variable. Therefore, it is seen that there is an
almost good correlation between the quantum patterns shown in figure
\ref{fig1} and classical trajectories. However, in order to show this correlation in a more clear sense, one may calculate the time dependence
of the expectation values. As we have seen in the second section, the variable $T$ can be regarded as an internal clock for the dynamics. The effective (reduced) Hamiltonian for this system can be found by solving the constraint ${\cal H}=0$ for $\pm P_T$ \cite{14}. Hence
\begin{equation}\label{rev1}
H_{eff}=-\frac{1}{6}a^{3\alpha-2}P_aP_{\phi}+\frac{1}{6}a^{3\alpha-3}\phi P_{\phi}^2-6ka^{3\alpha+1}\phi+a^{3\alpha+3}V(\phi). \end{equation}
With respect to this Hamiltonian the time evolution of the expectation value of an observable ${\cal O}$ reads as $\frac{d<{\cal O}>}{dT}=-i<\left[{\cal O},H_{eff}\right]>$. Therefore, we get
\begin{eqnarray}\label{rev2}
\left\{
\begin{array}{ll}
\frac{d<a>}{dT}=-\frac{1}{6}<a^{3\alpha-2}P_{\phi}>,\\\\
\frac{d<P_a>}{dT}=-\frac{1}{3}<a^{3\alpha-3}P_a P_{\phi}>+\frac{1}{2}<\phi a^{3\alpha-4}P_{\phi}^2>+6 k< a^{3\alpha}\phi>\\-3<a^{3\alpha +2}V(\phi)>+3\alpha P_0 <a^{-1}>,\\\\
\frac{d<\phi>}{dT}=-\frac{1}{6}<a^{3\alpha-2}P_{a}>+\frac{1}{3}<a^{3\alpha-3}\phi P_{\phi}>,\\\\
\frac{d<P_{\phi}>}{dT}=-\frac{1}{6}<a^{3\alpha-3}P_{\phi}^2>+6k<a^{3\alpha+1}>-<a^{3\alpha+3}V'(\phi)>.
\end{array}
\right.
\end{eqnarray}In the case of a very localized wave packet as we have constructed in figure \ref{fig1} (see also figure \ref{fig3} in the next subsection) a function $\Omega({\cal O},P)$ varies very little over the size of the wave packet and thus the approximation $<\Omega({\cal O},P)>\sim \Omega(<{\cal O}>,<P>)$ holds. Under this condition the trajectories for the expectation values obtained from the system (\ref{rev2}) are in agreement with the classical ones obtained from the system (\ref{O}) and shown in figures \ref{fig2}, \ref{fig4}. To summarize, in the present case where we have localized wave packets the expectation value of the scale factor is very close to the classical one and thus $<a>$ follows the pattern of $a(t)$ shown in figure \ref{fig2} with a good approximation.
\subsection{Radiation: $\alpha=1/3$}
In the standard Big-Bang model, it is believed that the early
universe is dominated either by radiation or radiation and high
energy particles. Since in this era the temperature is above
electron-positron pair threshold, these particles are relativistic
and the whole mixture behaves more like radiation rather than matter. When
the cosmological fluid is dominated by radiation, as it was presumably
the case in the early universe, the equation of state can be taken
as $p=\rho/3$. In this case, the solutions of equation (\ref{X}) may
be again separated  into the form $\psi(x,y)=X(x)Y(y)$ leading to
\begin{eqnarray}\label{AG}
\left\{
\begin{array}{ll}
\left[x^2\frac{d^2}{dx^2}-x\frac{d}{dx}+\left(1-4\nu^2-144kx^4+6x^6-24Ex^2\right)\right]X(x)=0,\\\\
\left[y^2\frac{d^2}{dy^2}+2y\frac{d}{dy}+\nu^2-\frac{1}{4}\right]Y(y)=0,
\end{array}
\right.
\end{eqnarray}
where we take $1-4\nu^2$ as a separation constant. Similar to the
previous subsection, we deal again with the solutions of the above
system in the case when $k=0$. In general, although the second
equation of the system (\ref{AG}) has exact solutions, its first
equation cannot be solved analytically. However, as we mentioned
above, the radiation is not the dominant fluid at the present epoch
but was dominant in early universe, i.e., in the regime where the scale factor is expected to have a small value. On the other hand, our classical analysis in this case (see below and figure \ref{fig4}) will show that the Ricci scalar follows a vanishing value. Therefore, we may write $x^2=a^2(2R)\sim\xi \ll1$ and hence $x^6=a^6(2R)^3\sim \xi a^4(2R)^2=\xi x^4$. Thus, we use the approximation
$x^6\sim \xi x^4$ in this era and rewrite the SWD equation as \footnote{Notice that this approximation leads to $R \sim \frac{\xi}{2a^2}$ which for a constant $a$ and very small value for $\xi$ goes to zero. The same result is also achieved from the definition of $R$ in (\ref{F}) when the scale factor is constant.}
\begin{eqnarray}\label{AH}
\left\{
\begin{array}{ll}
\left[x^2\frac{d^2}{dx^2}-x\frac{d}{dx}+\left(1-4\nu^2+6\xi x^4-24Ex^2\right)\right]X(x)=0,\\\\
\left[y^2\frac{d^2}{dy^2}+2y\frac{d}{dy}+\frac{1}{4}-\nu^2\right]Y(y)=0.
\end{array}
\right.
\end{eqnarray}
Now, the first equation of the system  has well known
solutions in terms of confluent hypergeometric functions $M(a,b;x)$ and $U(a,b;x)$. Therefore, we get
\begin{eqnarray}\label{AI}
X(x)&=&(i\sqrt{6\xi}x^2)^{\nu +1/2}e^{-i\frac{\sqrt{6\xi}}{2}x^2}\nonumber \\ && \times \left[C_1U\left(\nu+\frac{1}{2}-i\sqrt{6}E,2\nu+1;i\sqrt{6\xi}x^2\right)+C_2M\left(\nu+\frac{1}{2}-i\sqrt{6}E,2\nu+1;i\sqrt{6\xi}x^2\right)\right],\end{eqnarray}
\begin{equation}\label{AJ}
Y(y)=D_1y^{\nu-1/2}+D_2y^{-\nu-1/2},
\end{equation}
where $C_i$ and $D_i$ ($i=1,2$) are integration constants. Thus, the
SWD equation for radiation fluid has the following eigenfunctions
\begin{equation}\label{AK}
\Psi_{\nu E}(x,y,T)=e^{-iET}x^{2\nu+1}e^{-i\frac{\sqrt{6\xi}}{2}x^2}M\left(\nu+\frac{1}{2}-i\sqrt{6}E,2\nu+1;i\sqrt{6\xi}x^2\right)y^{-\nu-1/2},
\end{equation}
where we take $C_1=D_1=0$ to have well-defined eigenfunctions for
constructing the wave packets. To construct such wave packets since
the weight functions can be chosen as a shifted Gaussian function,
the general solutions to the SWD equation read as
\begin{equation}\label{AL}
\Psi(x,y,T)=\int_{\nu=0}^{\infty}\int_{E=0}^{\infty}e^{-\sigma(\nu-\nu_0)^2}e^{-\varsigma(E-E_0)^2}\Psi_{\nu E}(x,y,T)d\nu dE.
\end{equation}
\begin{figure}
\begin{tabular}{cccc} \epsfig{figure=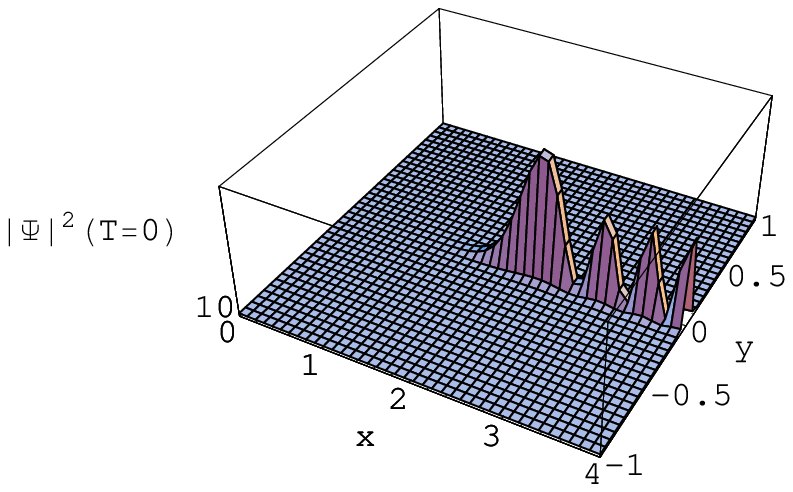,width=7cm}
\hspace{1cm} \epsfig{figure=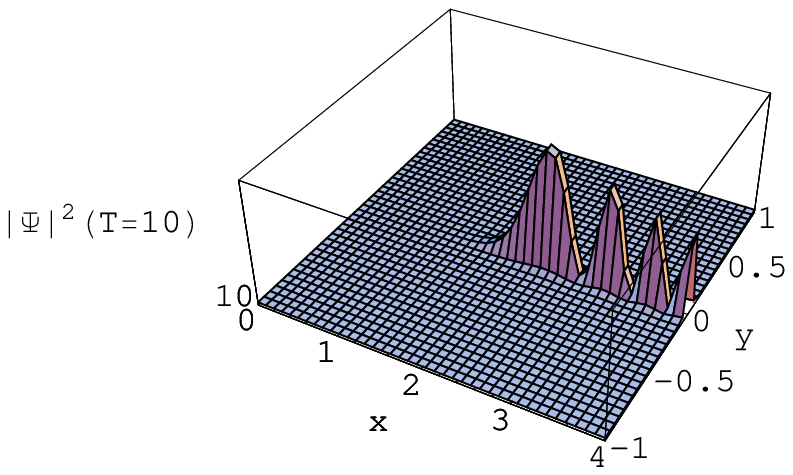,width=7cm}
\end{tabular}
\caption{\footnotesize The figures show
$|\Psi(x,y,T)|^2$, the square of the wavefunction in two different time parameter $T=0,10$. For larger $T$'s the same pattern is repeated.
The figures are plotted for the numerical values $\xi=0.1$, $\sigma=\varsigma=10$ and  $\nu_0=E_0=1$, and we have taken
the integrals in (\ref{AL}) from $0$ to $15$ for $E$ and from $0$ t0 $4$ for $\nu$.}
\label{fig3}
\end{figure}
In figure \ref{fig3} we have plotted the square of the above wavefunction for some numerical values of the parameters. The figures show that the wavefunction has several peaks which are sharply distributed around $y=0$. Also, it is seen that as time grows, the pattern of the wave packets do not show a major change. This means that in the context of our quadratic model the universe in its radiation era has almost a constant scale factor and thus its scalar curvature vanishes.

Now, let us pursue the corresponding classical cosmology given by
the system (\ref{O}) with $\alpha=1/3$, $V(\phi)=\frac{1}{4}\phi^2$
and $k=0$. In figure \ref{fig4} we have prepared the approximate
behavior of $a(t)$, and $\phi(t)\sim R(t)$ for typical values of the
parameters and initial conditions respectively. As the figure shows,
in complete agreement with the quantum model, the universe follows a
zero Ricci scalar while its scale factor has a constant value in the
epoch of the radiation dominated era of cosmic evolution.
\begin{figure}
\begin{tabular}{cccc} \epsfig{figure=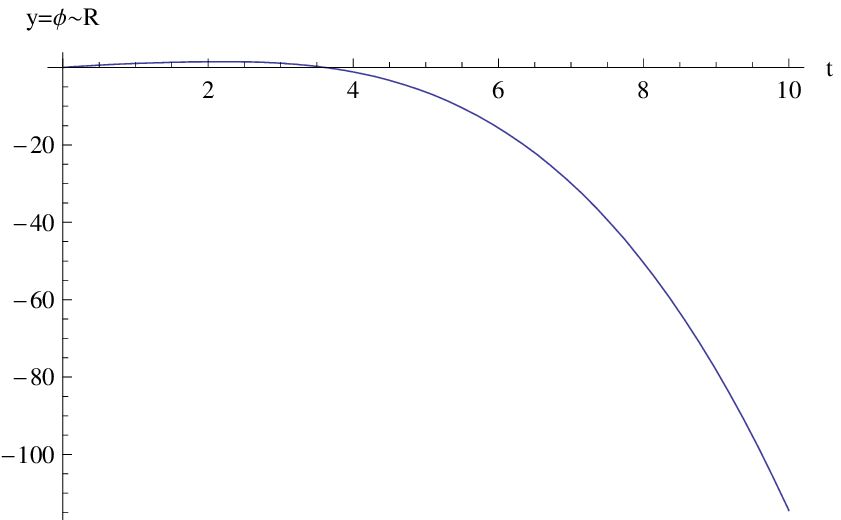,width=7cm}
\hspace{1cm} \epsfig{figure=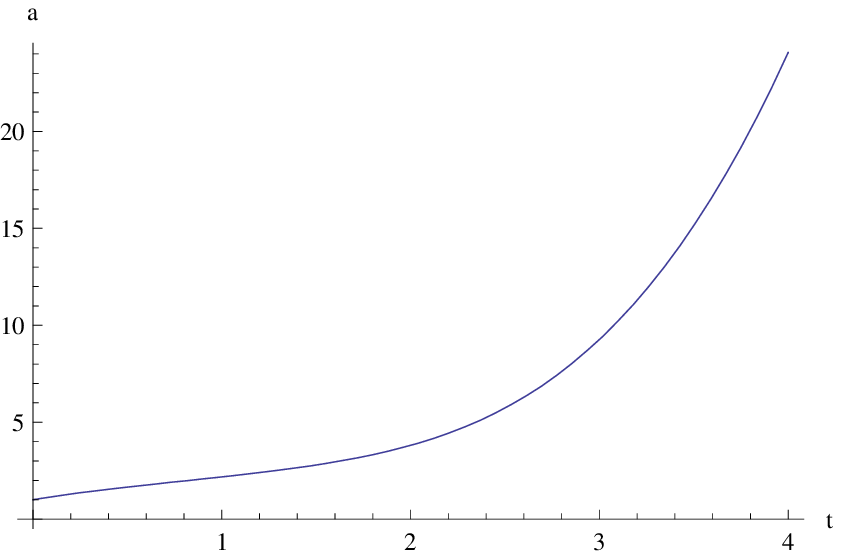,width=7cm}
\end{tabular}
\caption{\footnotesize Approximate behavior of $a(t)$, the scale
factor and $R(t)$, the Ricci scalar, in the classical cosmology
framework. The figures are plotted for numerical values
$\alpha=1/3$, $k= 0$ and $P_0=1$. We take the initial conditions
$a(t=0)=1$, $\phi(t=0)=0$, $P_a(t=0)=-10$ and $P_{\phi}(t=0)=-1$. After examining some other acceptable sets of initial conditions, we verify that this behavior is almost repeated.}
\label{fig4}
\end{figure}
\section{Comparison with $f(R)=R$ model}
In this section we briefly review the problem in the usual $f(R)=R$
framework which originally appeared in \cite{8}. The
Einstein-Hilbert Hamiltonian of the model used in the present work
can be written as
\begin{equation}\label{AM}
H=N{\cal H}= N\left(-\frac{P_a^2}{24 a}-6ka+\frac{P_T}{a^{3\alpha}}\right).\end{equation} Therefore, the equations of motion become
\begin{eqnarray}\label{AN}
\left\{
\begin{array}{ll}
\dot{a}=\{a,H\}=-\frac{N P_a}{12a},\\\\
\dot{P_a}=\{P_a,H\}=-N\left(\frac{P_a^2}{24a^2}-6k-3\alpha \frac{P_T}{a^{3\alpha +1}}\right),\\\\
\dot{T}=\{T,H\}=\frac{N}{a^{3\alpha}},\\\\
\dot{P_T}=\{P_T,H\}=0.
\end{array}
\right.
\end{eqnarray}Choosing the gauge $N=a^{3\alpha}$, we have $T=t$ and the following constraint equation ${\cal H}=0$
\begin{equation}\label{AO}
-6\frac{\dot{a}^2}{a^{3\alpha-1}}-6ka^{3\alpha+1}+P_0=0,\end{equation}where $P_T=P_0=\mbox{const.}$ as before. For flat space-time ($k=0$), the above equation has the solution
\begin{equation}\label{AP}
a(t)=a_0e^{\sqrt{\frac{P_0}{6}}t},\end{equation}for stiff matter ($\alpha=1$), and
\begin{equation}\label{AQ}
a(t)=\sqrt{\frac{P_0}{6}}t+a_0,\end{equation} for radiation
($\alpha=1/3$)\footnote{In terms of cosmic time $\eta=\int Ndt$,
these solutions read as $a(\eta)\sim \eta^{1/3}$ for stiff matter
and $a(\eta)\sim \eta^{1/2}$ for radiation.}. It is clear that for
both of these solutions the Ricci scalar
$R=\frac{6}{N^2}\left(\frac{\ddot{a}}{a}+\frac{\dot{a}^2}{a^2}-\frac{\dot{N}\dot{a}}{Na}\right)$
has a behavior $R\rightarrow 0$ as $t\rightarrow +\infty$. In the
case when the universe is filled with stiff matter, comparison of
the solution (\ref{AP}) with those of the quadratic model which
appear in figure \ref{fig2}, shows that while the scale factors have
the same behavior at late times, their early-time evolution does not coincide with each other. This means that although the quadratic
model can predict the late time expansion of the universe, but in
spite of the usual $f(R)=R$ model its Ricci scalar grows
monotonically with time. Also, for radiation the results of two
models show some differences. In $f(R)=R$ cosmology the scale factor
has a linearly expanding behavior according to (\ref{AQ}), with a
decreasing Ricci scalar while the quadratic cosmology shows a static
universe with zero Ricci scalar. To give an estimate why the result of the quadratic $R^2$ -cosmology decline from those of the linear Einstein-Hilbert theory, we should note that although this form of $f(R)$ gravity yields a late time expansion cosmology, but in view of having the correct weak-field limit at Newtonian and post-Newtonian levels has not a desired form. The conditions under which a modified
gravity model passes the local and astrophysical tests such as Newton law and
solar system tests are investigated in \cite{15}. In these works such $f(R)$ theories are
studied which satisfy the conditions
\begin{equation}\label{AR}
\lim_{R\rightarrow \infty} f(R)=\mbox{cons.},\hspace{.5cm}\lim _{R\rightarrow 0}f(R)=0,\end{equation}and shown that they pass Newton law, stability of Earth-like gravitational solution, heavy mass for additional scalar degree of freedom, etc. Therefore, since our quadratic model does not satisfy the above conditions, it is not a viable theory with correct Newtonian and post-Newtonian limits. This is not surprising
since it is well known that a large class of $f(R)$ theories suffer from this
issue \cite{CH}. In summary, in the framework of
quadratic model, although we have a compatible classical and quantum
cosmology discussed in previous sections, the predictions of this
model do not coincide with the usual $f(R)=R$ case.
\section{Conclusions}
In this paper we have studied a quantum cosmological model in the
framework of quadratic $f(R)$ gravity coupled to the Schutz' perfect
fluid. The use of Schutz' formalism for perfect fluid allowed us to
obtain a SWD equation in which the only remaining matter degree of
freedom plays the role of time parameter in the model. For stiff
fluid ($\alpha=1$) and radiation ($\alpha=1/3$), we saw that the SWD
equation can be separated and its eigenfunctions can be obtained in
terms of known special functions. In the
case of stiff matter as the cosmic fluid, we found the eigenstates
in terms of Bessel functions and then constructed the corresponding
wave packets by appropriate superposition of the eigenstates. The
wavefunction in this case shows a pattern in which there are
different possible quantum states (with different probability) from
which our present universe could have evolved and tunneled in the
past from one state to another. The time evolution of this wave
packet represents its motion along the larger $R$ direction. We have
also solved the corresponding classical cosmology in this case by
numerical methods and showed that the Ricci scalar has a
monotonically growing behavior with time while the scale factor
first decreases and then increases forever. Therefore, we have a
model with compatible classical and quantum cosmology which although
can predict the late time expansion of the universe, but contrary to
the usual standard model of cosmology the universe will admit states
with larger $R$ in its late time evolution. On the other hand, in
the case of radiation, we obtained wavefunctions which preserve
their configuration during time evolution. These wavefunctions peak
sharply around the zero Ricci scalar. The classical cosmology in
this case is again in agreement with the quantum model and shows an
almost static universe with zero scalar curvature. Since in the
standard model the universe has an expansion behavior as
$a(\eta)\sim \eta^{1/2}$ ($\eta$ is the cosmic time) in its
radiation dominated era, our quadratic model does not coincide with
real cosmological observations also in this epoch.
\vspace{5mm}\newline \noindent {\bf
Acknowledgements}\vspace{2mm}\noindent\newline
The author is grateful to the anonymous referees
for enlightening suggestions, to H. R. Sepangi for a careful reading of the manuscript and helpful comments
and to the research council of Azad University of Chalous for
financial support.


\begin{thebibliography}{99}
\bibitem{1} C. Deffayet, {\it Phys. Lett.} B {\bf 502} (2001) 199
(arXiv: hep-th/0010186)\\J.S. Alcaniz, {\it Phys. Rev.} D {\bf 65}
(2002) 123514 (arXiv: astro-ph/0202492)\\S.M. Carroll, V. Duvvuri,
M. Trodden and M. Turner, {\it Phys. Rev.} D {\bf 70} (2004) 043528
(arXiv: astro-ph/0306438)\\S. Nojiri and S.D. Odintsov, {\it Phys.
Lett.} B {\bf 576} (2003) 5 (arXiv: hep-th/0307071)\\K. Atazadeh and
H.R. Sepangi, {\it Int. J. Mod. Phys.} D {\bf 16} (2007) 687
(arXiv: gr-qc/0602028)\\S. Nojiri, S.D. Odintsov and M. Sami, {\it
Phys. Rev.} D {\bf 74} (2006) 046004 (arXiv: hep-th/0605039)\\S. Capozziello and M. Francaviglia, {\it Gen. Rel. Grav.} {\bf 40} (2008) 357 (arXiv: 0706.1146 [gr-qc])\\S. Capozziello, {\it
Int. J. Mod. Phys.} D {\bf 11} (2002) 483 (arXiv: gr-qc/0201033)\\S. Nojiri and S. D. Odintsov, {\it Phys.Rev.} D {\bf 68} (2003) 123512 (arXiv:hep-th/0307288)\\ S.K. Chakrabarti, E.N. Saridakis and A.A. Sen,
{\it A new approach to modified-gravity models} (arXiv: 0908.0293 [astro-ph.CO])

\bibitem{2}S. Nojiri and S.D. Odintsov, {\it Dark energy, inflation and dark matter from modified $f(R)$
gravity} (arXiv: 0807.0685 [hep-th])\\S. Nojiri and S.D. Odintsov, {\it Int. J. Geom. Methods Mod. Phys.} {\bf 4} (2007) 115 (arXiv: hep-th/
0601213)\\ T.P. Sotiriou and V. Faraoni, {\it $f(R)$ theories of gravity} (arXiv: 0805.1726 [gr-qc])

\bibitem{rev1} D. Lovelock, {\it J. Math. Phys.} {\bf 12} (1971) 498

\bibitem{rev2}K.S. Stelle, {\it Phys. Rev.} D {\bf 16} (1977) 953

\bibitem{rev3} R. Utiyama and B. DeWitt, {\it J. Math. Phys.} {\bf 3} (1962) 608

\bibitem{rev4} K.S. Stelle, {\it Gen. Rel. Grav.} {\bf 9} (1978) 353\\G.T. Horowitz and R.M. Wald, {\it Phys. Rev.} D {\bf 17} (1978) 414

\bibitem{rev5} K. Kleidis, A. Kuiroukidis and D.B. Papadopoulos, {\it Phys. Lett.} B {\bf 546} (2002) 112

\bibitem{3} D.N. Vollick, {\it Phys. Rev.} D {\bf 68} (2003)
063510 (arXiv: astro-ph/0306630)\\X. Meng and P. Wang, {\it Class.
Quantum Grav.} {\bf 20} (2003) 4949 (arXiv: astro-ph/0307354)\\X.
Meng and P. Wang, {\it  Palatini formulation of modified gravity
with squared scalar curvature  } (arXiv: astro-ph/0308284)\\ X. Meng and P. Wang, {\it Phys. Lett.} B {\bf 584} (2004) 1 (arXiv: hep-th/0309062)

\bibitem{CH}T. Chiba, {\it Phy. Lett.} B {\bf 575} (2003) 1 (arXiv: astro-ph/0307338)\\
E.E. Flanagan, {\it Class. Quantum Grav.} {\bf 21} (2004) 3817 (arXiv: gr-qc/0403063)\\
T.P. Sotiriou, {\it Class. Quantum Grav.} {\bf 23} (2006) 5117 (arXiv: gr-qc/0604028)

\bibitem{4}T.P. Sotiriou and S. Liberati, {\it Ann. Phys.} {\bf
322} (2007) 935 (arXiv: gr-qc/0604006)\\N.J. Poplawski, {\it Class.
Quantum Grav.} {\bf 23} (2006) 2011 (arXiv: gr-qc/0510007)

\bibitem{5}M.C.B. Abdalla, A. Nojiri and S.D. Odintsov, {\it Class. Quantum Grav.} {\bf 22} (2005) L35 (arXiv: hep-th/0409177)
\\X. Meng and P. Wang,{\it Class. Quantum Grav.} {\bf 22} (2005) 23 ( arXiv: gr-qc/0411007)\\X. Meng and P. Wang, {\it Class. Quantum
Grav.} {\bf 21} (2004) 951 (arXiv: astro-ph/0308031)\\T. Clifton and J.D. Barrow, {\it Phys. Rev.} D {\bf 72} (2005) 103005 (arXiv: gr-qc/0509059)\\T. Clifton, {\it Class. Quantum Grav.} {\bf 23}
(2006) 7445 ( arXiv: gr-qc/0607096)\\ K. Atazadeh and H.R. Sepangi, {\it Phys. Lett.} B {\bf 643} (2006) 76 (arXiv: gr-qc/0610107)\\J.D. Barrow and T. Clifton, {\it Class.Quantum Grav.} {\bf 23} (2006) L1 ( arXiv: gr-qc/0509085)\\B. Vakili, {\it Phys. Lett.} B {\bf 664} (2008) 16 (arXiv: 0804.3449 [gr-qc])\\ K. Atazadeh, M. Farhoudi and H.R. Sepangi, {\it Phys. Lett.} B {\bf 660} (2008) 275 (arXiv: 0801.1398 [gr-qc])

\bibitem{6}A. Shojai and F. Shojai, {\it Gen. Rel. Grav.} {\bf 40} (2008) 1967 (arXiv: 0801.3496 [gr-qc])\\B. Vakili, {\it Phys. Lett.} B
{\bf 669} (2008) 206 (arXiv: 0809.4591 [gr-qc])\\G. Cognola, E. Elizalde, S. Nojiri, S.D. Odintsov and S. Zerbini, {\it J. Cosmol. Astropart. Phys.} {\bf 0502} (2005) 010 (arXiv: hep-th/0501096)\\A.K. Sanyal and
B. Modak, {\it Phys. Rev.} D {\bf 63} (2001) 064021 (arXiv: gr-qc/0107001)\\
A.K. Sanyal and B. Modak, {\it  Class.Quantum Grav.} {\bf 19} (2002) 515 (arXiv: gr-qc/0107070)\\M. Kenmoku, K. Otsuki, K. Shigemoto and K. Uehara, {\it  Class. Quantum Grav.} {\bf 13} (1996) 1751 (arXiv: gr-qc/9510027)\\U. Kasper, {\it Class. Quantum Grav.} {\bf 10} (1993) 869\\L.O. Pimentel, O. Obregon and J.J. Rosales, {\it Class. Quantum Grav.} {\bf 14} (1997) 379\\L.O. Pimentel and O. Obregon, {\it Class. Quantum Grav.} {\bf 11} (1994) 2219

\bibitem{7} B.F. Schutz, {\it Phys. Rev.} D {\bf 2} (1970) 2762\\B.F. Schutz, {\it Phys. Rev.} D {\bf 4} (1971) 3559

\bibitem{8}A.B. Batista, J.C. Fabris, S.V.B. Goncalves and J. Tossa, {\it Phys. Lett.} A {\bf 283} (2001) 62
(arXiv: gr-qc/0011102)\\F.G. Alvarenga, J.C. Fabris, N.A. Lemos and
G.A. Monerat, {\it Gen. Rel. Grav.} {\bf 34} (2002) 651 (arXiv: gr-qc/0106051)\\A.B. Batista, J.C. Fabris, S.V.B. Goncalves and J. Tossa, {\it Phys. Rev.} D {\bf 65} (2002) 063519 (arXiv: gr-qc/0108053)\\N.A. Lemos, {\it J. Math. Phys.} {\bf 37} (1996) 1449 (arXiv: gr-qc/9511082)\\P. Pedram, S. Jalalzadeh and S.S. Gousheh, {\it Class. Quantum Grav.} {\bf 24} (2007) 5515 (arXiv: 0709.1620 [gr-qc])\\P. Pedram and S. Jalalzadeh, {\it Phys. Rev.} D {\bf 77} (2008) 123529 (arXiv: 0805.4099 [gr-qc])

\bibitem{DW} B.S. DeWitt, {\it Phys. Rev.} {\bf 160} (1967) 1113

\bibitem{9}V.G. Lapchinskii and V.A. Rubakov, {\it Theor. Math. Phys.} {\bf 33} (1977) 1076

\bibitem{10}A. Anderson, {\it Ann. Phys.} {\bf 232} (1994) 292 (arXiv: hep-th/9305054)

\bibitem{11} S.W. Hawking and D.N. Page, {\it Nucl. Phys.} B {\bf 264} (1986) 185

\bibitem{12}R. Steigl and F. Hinterleitner, {\it Class. Quantum Grav.} {\bf 23} (2006) 3879

\bibitem{13}M. Abramowitz and I.A. Stegun, {\it Handbook of Mathematical Functions} (1972) (New York: Dover)

\bibitem{14} C.J. Isham, {\it Canonical quantum gravity and the problem of time} (arXiv: gr-qc/9210011)
\\B. Vakili and H.R. Sepangi, {\it Ann. Phys.} {\bf 323} (2008) 548 (arXiv: 0709.2988 [gr-qc])

\bibitem{15} W. Hu and I. Sawicki, {\it Phys. Rev.} D {\bf 76} (2007) 064004
(arXiv: 0705.1158 [astro-ph])\\S. Nojiri and S.D. Odintsov, {\it Newton law corrections and instabilities in $f(R)$ gravity with the effective cosmological constant epoch} (arXiv: 0706.1378 [hep-th])\\S. Nojiri and S.D. Odintsov, {\it Phys. Rev.} D {\bf 77} (2008) 026007
(arXiv: 0710.1738 [hep-th])
\end{thebibliography}
\end{document}